\documentclass[twocolumn,aps,amsmath,amssymb]{revtex4}

\def\al{\alpha}
\def\be{\beta}

\def\bfa{{\boldsymbol{\alpha}}}

\def\cI{{\cal I}}

\def\bi{{\bf a}}
\def\bj{{\bf b}}
\def\bk{{\bf c}}

\def\bE{{\bf 1}}
\def\bZ{{\bf 2}}
\def\bD{{\bf 3}}
\def\bV{{\bf 4}}
\def\bS{{\bf 6}}
\def\bA{{\bf 8}}

\def\bbE{{\bar{\bf 1}}}
\def\bbZ{{\bar{\bf 2}}}
\def\bbD{{\bar{\bf 3}}}
\def\bbV{{\bar{\bf 4}}}
\def\bbS{{\bar{\bf 6}}}

\def\bbi{{\bar{\bf a}}}
\def\bbj{{\bar{\bf b}}}
\def\bbk{{\bar{\bf c}}}

\def\11{{\mathbb 1}}

\def\beq{\begin{equation}}
\def\eeq{\end{equation}}
\def\bea{\begin{eqnarray}}
\def\eea{\end{eqnarray}}
\def\nn{\nonumber}

\def\ra{\rightarrow}

\begin{document}
\draft
\title{Standard Model Fermions and N=8 supergravity}
\author{Krzysztof A. Meissner$^{1,2}$ and Hermann Nicolai$^1$}
\affiliation{$^1$ Max-Planck-Institut f\"ur Gravitationsphysik
(Albert-Einstein-Institut)\\
M\"uhlenberg 1, D-14476 Potsdam, Germany\\
$^2$ Institute of Theoretical Physics, Faculty of Physics,
University of Warsaw\\
Pasteura 5, 02-093 Warsaw, Poland}

\vspace{3mm}

\begin{abstract} In a scheme originally proposed by M. Gell-Mann, and subsequently
shown to be realized at the SU(3)$\,\times\,$U(1) stationary point of maximal gauged
SO(8) supergravity by N.~Warner and one of the present authors, the 48 spin-$\frac12$
fermions of the theory remaining after the removal of eight Goldstinos can be identified
with the 48 quarks and leptons (including right-chiral neutrinos) of the
Standard Model, provided one identifies the residual SU(3) with the diagonal
subgroup of the color group SU(3)$_c$ and a family symmetry SU(3)$_f$.
However, there remained a systematic mismatch in the electric charges by a spurion
charge of $\pm \frac16$. We here identify the `missing'  U(1) that rectifies
this mismatch, and that takes a surprisingly simple, though unexpected form.
\end{abstract}
]
\maketitle

%\section{Introduction}

Maximal gauged $N\!=\!8$ supergravity \cite{dWN} admits six AdS vacua (critical points)
at which the SO(8) symmetry is broken to a subgroup containing SU(3) \cite{Warner}.
Of these, the one with unbroken SU(3)$\,\times\,$U(1) symmetry is in several ways
the most interesting \cite{NW}. In addition to the residual gauge symmetry, it preserves
$N\!=\!2$ supersymmetry, such that its properties can be fully analyzed by means
of $N\!=\!2$ AdS supermultiplets \cite{CFN,NW}. Furthermore,  the group SU(3)$_c\,\times\,$U(1)$_{em}$ is the gauge symmetry that survives to the lowest energies in the
Standard Model. However, a naive identification of the supergravity SU(3) with
the color group SU(3)$_c$ does not work, as is immediately obvious from
the decompositions displayed below (cf. eq.~(\ref{Fermions1})). For this reason,
M.~Gell-Mann introduced an additional family symmetry SU(3)$_f$ that acts
between the three particle families (generations) and proposed to identify
the residual SU(3) of supergravity with the {\em diagonal subgroup}
of color and family \cite{GellMann}. This scheme `almost'  works in the
sense that, after the removal of eight Goldstinos (as required for a {\em complete} breaking
of supersymmetry)  there is complete agreement of the the SU(3) assignments,
but there remains a systematic mismatch between the U(1) charges: the electric charges
of the supergravity fermions are systemically off by $\pm \frac16$ from those of
the quarks and leptons. Nevertheless, and especially  in view of the persistent failure by LHC
to detect any new fundamental spin-$\frac12$ degrees of freedom (so we `may have already
seen it all'), the agreement between the observed number of quarks and leptons, and
the number of physical spin-$\frac12$ fermions in maximal supergravity remaining
after complete breaking of supersymmetry is a tantalizing coincidence \cite{fn}.

In this article we identify the `missing' U(1) symmetry (designated by U(1)$_q$)
that rectifies the mismatch in the electric charge assignments. As it turns out its action
on the original 56 fermions is surprisingly simple, but requires a `deformation' of the residual
SU(3)$\,\times\,$U(1) symmetry reminiscent of the deformation that appears in
non-trivial co-products. We do not know whether and how such a deformation could be realized dynamically, but the final result (see (\ref{cI}) below) is of such a suggestive simplicity that  we may take it as a hint of some non-trivial underlying dynamics that could also lead to new ways of dynamically breaking supersymmetry, possibly in a framework beyond maximal supergravity. Consequently, one main
message here is that the `linear'  decompositions of group representations commonly
employed (often in cascade-like sequences of symmetry breakings)
to obtain the particle content of the low energy theory may not suffice to explain
the emergence of the Standard Model from a unified Planck scale theory.

Let us begin by briefly recalling some basic properties of $N\!=\!8$ supergravity.
In its original ungauged version \cite{CJ} the theory possesses a linearly realized
global E$_{7(7)}$ symmetry and a local chiral SU(8) symmetry, with composite
SU(8) gauge fields. Upon choosing a special SU(8) gauge the local SU(8) symmetry
collapses to a global (or `rigid') SU(8); in this gauge the non-compact part of E$_{7(7)}$ is realized non-linearly. There is no potential for the scalar fields (`moduli'), hence there remains a large vacuum degeneracy. This degeneracy is lifted by gauging the theory. To this aim one
promotes an SO(8) subgroup of E$_{7(7)}$ to a local symmetry, with the 28 spin-1 fields
of $N\!=\!8$ supergravity as the Yang-Mills vector bosons \cite{dWN} (thanks to modern
techniques based on the embedding tensor there now exists a large variety of
other gaugings, see e.g \cite{dWST,DAIT}, but the SO(8) gauging remains the only
one with a compact simple gauge group). To maintain full local supersymmetry,
the Lagrangian must be modified by Yukawa couplings and a scalar potential, which
has been found to display a wealth of stationary points (see Ref.\cite{Fischbacher},
which lists 41 extrema, and Ref.\cite{CDIZ} for a more recent survey that also discusses
other gaugings). Properties of the SU(3)$\,\times\,$U(1) stationary point are discussed
at length in \cite{NW}, to which we refer for further details. Let us also note that
the group theoretical decompositions presented here are independent of the
dynamics, and thus to some extent also independent of the specifics of the
stationary point. They could thus also apply to some of the  SU(3)$\,\times\,$U(1) extrema
of the new gaugings recently studied in \cite{CDIZ}.

In the remainder we focus on the fermionic sector of the theory, which
consists of eight gravitinos $\psi_\mu^i$
transforming in the $\bA$, and a tri-spinor of spin-$\frac12$ fermions $\chi^{ijk}$
transforming in the $\bf{56}$ of SU(8), whence  $\chi^{ijk}$ is fully
antisymmetric in the SU(8) indices $i,j,k$. We here follow the conventions and
notations of \cite{dWN}, so complex conjugation raises (or lowers) indices, such that
for instance $\chi^{ijk} = \big( \chi_{ijk} \big)^*$; at the same time upper (lower)  position
of the SU(8) indices indicates positive (negative) chirality.
%(this is essentially  the same as Weyl spinor notation with dotted and undotted spinors
%$\chi^{ijk}_\al$ and $\chi_{ijk\dot\al}$, respectively).
Hence the chiral SU(8) transformations act as
\beq\label{Uchi}
\chi^{ijk} \; \ra \; U^i{}_l U^j{}_m U^k{}_n \chi^{lmn} \; , \;\;
\chi_{ijk} \; \ra \; U_i{}^l U_j{}^m U_k{}^n \chi_{lmn}
\eeq
with $U \in$ SU(8), and $U_i{}^j \equiv \big( U^i{}_j \big)^*$, whence the
unitarity relation  $U^\dagger U = \bE$ is equivalently expressed by
$U^i{}_k U_j{}^k = \delta^i_j$. When a special SU(8) gauge chosen,
the remaining local SO(8) acts by real orthogonal transformations $O^i{}_j$,
and thus no longer chirally on the fermions.

The group SO(8) admits a subgroup U(3)$\,\times\,$U(1), via
the embedding SO(6)$\,\times\,$SO(2) $\subset$ SO(8). To study the relevant
decompositions we introduce boldface indices and their complex conjugates
according to  \cite{NW}
\bea
V^\bE &\equiv& V^1 + i V^2 \; , \quad   V^\bbE = V^1 -i V^2, \quad\nn\\
V^\bZ &\equiv& V^3 + i V^4 \; , \quad   V^\bbZ = V^3 -i V^4,   \nn\\
V^\bD &\equiv& V^5 + i V^6 \; , \quad   V^\bbD \equiv V^5 -i V^6, \quad\nn\\
V^\bV &\equiv& V^7 + i V^8 \; , \quad   V^\bbV \equiv V^7 -i V^8    \nn
\eea
so the complex conjugate representations are indicated by putting a bar
on these indices. The U(3) acts on the first three indices $\bi ,\bj, \dots= \bE,\bZ,\bD$.
The boldface indices thus furnish a compact way of writing the SU(3) representations;
writing them out  in terms of the original SU(8) fermions $\chi^{ijk}$ we have, for instance,
\bea
\chi^{\bE\bZ\bbV} &=& \chi^{137} + i\chi^{237} + i\chi^{147} - i\chi^{138} \nn\\[1mm]
          &&-\chi^{247} + \chi^{238} + \chi^{148} + i \chi^{248}\  \nn\\[2mm]
\chi^{\bE\bbE\bV} &=& -2 i \chi^{127} + 2 \chi^{128}
\eea
and so on. The group U(1)$\,\times\,$U(1) is a two parameter abelian subgroup
whose associated Lie algebra is embedded as follows into SO(8)
\begin{equation} \label{U(1)}
 Y(\al,\be) \,= \, \begin{pmatrix}
                0 & \al  & 0 & 0 & 0 & 0 & 0 & 0 \\
                -\al  & 0 & 0 & 0 & 0 & 0 & 0 & 0 \\
                0 & 0 & 0 & \al & 0 & 0 & 0 & 0 \\
                0 & 0 & - \al & 0 & 0 & 0 & 0 & 0 \\
                0 & 0 & 0 & 0 & 0 & \al & 0 & 0 \\
                0 & 0 & 0 & 0 & - \al & 0 & 0 & 0 \\
                0 & 0 & 0 & 0 & 0 & 0 & 0 & \beta\\
                0 & 0 & 0 & 0 & 0 & 0 & - \beta & 0
                \end{pmatrix},
\end{equation}
This matrix commutes with U(3)$\,\times\,$U(1) $\subset$ SO(8) for all $\al, \be$.
Consequently, for each choice of $\al$ and $\be$ the above matrix defines
an SU(3)$\,\times\,$U(1) subgroup of SO(8) where we denote U(1) $\equiv$
U(1)$_{\al,\be}$ for simplicity.

Given some choice of $\al,\be$, one easily reads off the
the SU(3)$\,\times\,$U(1) assignments for the gravitinos
\bea\label{Gravitino1}
\psi_\mu^\bi &\in&  (\bD\,,\, \al)\; , \quad \psi_\mu^\bbi \in (\bbD\,,\, - \alpha)\; ,  \nn\\[2mm]
\psi_\mu^\bV &\in& (\bE \,,\,\beta) \; , \quad \psi_\mu^\bbV \in (\bE \,,\, - \beta)
\eea
The 56 spin-$\frac12$ fermions are split into six Goldstinos
\beq\label{Goldstino}
\chi^{\bi\bV\bbV} \,\in\, (\bD\,,\, \al) \; , \quad
\chi^{\bbi\bV\bbV} \,\in \, (\bbD\,,\, - \al) \;\; ,
\eeq
two `would-be Goldstinos'
\beq\label{PseudoGoldstino}
\chi^{\bi\bj\bk} \in (\bE\,,\, 3\al) \;,\quad \chi^{\bbi\bbj\bbk} \,\in \,(\bE\,,\, -3\al)
\eeq
and the remaining 48 spin-$\frac12$ fermions:
{\small
\begin{align}\label{Fermions1}
\chi^{\bi\bj\bV} &\in (\bbD\,,\, 2\al + \be) \; , \; &\chi^{\bi\bj\bbV} \,&\in\, (\bbD\,,\, 2\al - \be)
\nn\\[1mm]
\chi^{\bbi\bbj\bV} &\in (\bD\,,\, - 2\al + \be) \; , \; &\chi^{\bbi\bbj\bbV} \,&\in\, (\bD\,,\, -2\al- \be)
\\[1mm]
\chi^{\bi\bj\bbk} &\in (\bD,\al) \oplus (\bbS,\al) \;,\;
&\chi^{\bbi\bbj\bk} &\in (\bbD, - \al) \oplus (\bS\,,\,-\al) \;\;
\nn\\[1mm]
\chi^{\bi\bbj\bV} &\in ({\bf 8}\,,\,\be) \oplus (\bE\, ,\,\be) \;,\;
&\chi^{\bi\bbj\bbV} &\in ({\bf 8}\,,\, - \be) \oplus (\bE\,,\, -\be)  \nn
\end{align}
}
At the SU(3)$\,\times\,$U(1) stationary point \cite{Warner} the $N\!=\!8$ supersymmetry
is broken to $N\!=\!2$ supersymmetry, with two massless gravitinos
$\psi^\bV_\mu \equiv \psi_\mu^7 + i \psi_\mu^8$ and
$\psi_\mu^\bbV \equiv \psi_\mu^7 - i \psi_\mu^8$,
while the six Goldstinos (\ref{Goldstino}) are eaten to give six massive
gravitinos $\psi_\mu^\bi$ and $\psi_\mu^\bbi$. As shown in \cite{NW},
all particles fit properly into multiplets of $N\!=\!2$ AdS supersymmetry.
The mass eigenstates at the stationary point actually mix those fermions
lying in the same SU(3)$\,\times\,$U(1) representations (see \cite{NW} for explicit
formulas and a full analysis of the AdS mass spectrum), but these would anyhow
have to re-group along the `deformed' SU(3)$\,\times\,$U(1) to be presented below, if the latter is dynamically excited. Furthermore, in terms of the original chiral SU(8) we still have a residual chiral SU(2)  R-symmetry which, in terms of the
original SU(8) acts on the indices $i,j,\dots = 7,8$ and commutes with
the SU(3) factor.

To get agreement with the non-supersymmetric low energy world, the residual
$N=2$ supersymmetry must, of course, also be broken, and this must happen through some as yet unknown dynamical mechanism. In this last step the remaining massless
gravitinos $\psi_\mu^\bV$ and $\psi_\mu^\bbV$ would eat the `would-be
Goldstinos' from (\ref{PseudoGoldstino}) to become massive, whence we
are left with the fermions listed in (\ref{Fermions1}).  The challenge is then to match these remaining 48 spin-$\frac12$ fermions with those of the Standard Model.

Now, as shown in \cite{NW}, the residual $N\!=\!2$ supersymmetry and the structure of (long and short) multiplets of $N\!=\!2$ AdS supersymmetry \cite{CFN,NW} require
\beq\label{albe}
\al = \frac16 \;\; , \quad\be = \frac12 \, ,
\eeq
Remarkably, this choice is also the one required for the matching  with quarks and
leptons, modulo a spurion charge $q$ \cite{Min}. Namely, if -- besides the standard color
charge assignments -- we assign all fermions to triplets or anti-triplets of a new
family group SU(3)$_f$ in the way indicated below,  the identification
(after removing all eight Goldstinos) with quarks and leptons is  \cite{GellMann}
{\small
\begin{align}
\label{Fermions2}
\chi^{\bi\bbj\bV}\; &: \
(u,c,t)_L &   \, &\, \bD_c \times \bbD_f    \,\ra \, \bA \oplus \bE  & \
\frac23 &= \frac12 + q \nn\\
\chi^{\bbi\bj\bbV}\; & : \
(\bar{u}, \bar{c}, \bar{t})_L &  \, &\, \bbD_c \times \bD_f  \,\ra \, \bA \oplus \bE & \ - \frac23 & = -\frac12 - q\nn\\
\chi^{\bbi\bbj\bk}\; &: \
(d,s,b)_L & \, & \, \bD_c \times \bD_f   \, \ra \, \bS \oplus \bbD  &
\  - \frac13 & =  - \frac16 -  q \nn\\
\chi^{\bi\bj\bbk}\; &: \
(\bar{d}, \bar{s}, \bar{b})_L &  \,& \, \bbD_c \times \bbD_f  \, \ra \,\bbS \oplus \bD & \ \frac13 & = \frac16 + q\nn\\
\chi^{\bi\bj\bbV}\; &: \
(\nu_e,\nu_\mu,\nu_\tau)_L &  \, & \, \bE_c \times \bbD_f    \,\ra \,
\bbD  & \  0 &=  - \frac16 +  q \nn\\
\chi^{\bbi\bbj\bV}\; &: \
(\bar{\nu}_e,\bar{\nu}_\mu, \bar{\nu}_\tau)_L & \, & \, \bE_c \times \bD_f  \, \ra \, \bD & \
0 &= \frac16 - q\nn\\
\chi^{\bbi\bbj\bbV}\; &: \
(e^-,\mu^-,\tau^-)_L &  \, & \, \bE_c \times \bD_f    \,\ra \,
\bD  & \ - 1 & =  - \frac56  -  q \nn\\
\chi^{\bi\bj\bV}\; &: \
(e^+,\mu^+,\tau^+)_L &  \,  & \, \bE_c \times \bbD_f  \, \ra \, \bbD
& \ 1  & =  \frac56  +  q
\end{align}
}
where we made use of
the fact (well known to GUT practitioners) that right-chiral particles can be
equivalently described by their left-chiral anti-particles. The most important
feature here is that the SU(3) of $N=8$ supergravity is {\em not} identified with
the QCD color group SU(3)$_c$, but rather with the {\em diagonal subgroup}
of color and family symmetry, that is, we identify
\beq
SU(3) \equiv \big[SU(3)_c \times SU(3)_f \big]{}_{diag}.
\eeq
Breaking color and family symmetry to the diagonal subgroup may look strange,
but a not so dissimilar scheme does appear to work surprisingly well
in pure QCD with three flavors, if one assumes that the product of color
and flavor SU(3) symmetries is broken to the diagonal SU(3) subgroup
by a diquark condensate \cite{Wilczek} (`flavor-color locking').
The last column in (\ref{Fermions2}) shows the physical electric charges (on the left)
in comparison with the U(1) charges as obtained from the decomposition of
the $N=8$ fermions (on the right, that is (\ref{Fermions1}) with the particular choice (\ref{albe})). As we see, the latter differ from the quark and lepton charges systematically by the spurion charge $q$, with negative (positive) sign for family triplets (anti-triplets).
Accordingly  the spurion charge must be taken $q= \frac16$ to get agreement with the
electric charges of quarks and leptons \cite{GellMann}. Importantly,
the electroweak SU(2)$_w$  would {\em not} commute with SU(3)$_f$, as the upper and lower components of the would-be electroweak doublets are assigned to opposite representations of
SU(3)$_f$. More precisely, the upper components of the would-be electroweak
doublets [that is, $(u,c,t)_L$ and $(\nu_e, \nu_\mu, \nu_\tau)$] are assigned to
the $\bbD_f$ of SU(3)$_f$, while their lower components [that is, $(d,s,b)_L$
and $(e^-,\mu^-,\tau^-)_L)$] are assigned to the $\bD_f$ of SU(3)$_f$.
As a consequence, the residual chiral SU(2) R-symmetry at the stationary point
can {\em not} be identified with the electroweak SU(2)$_w$.

We now look for an implementation of the missing $q$-rotation on the
56 spin-$\frac12$ fermions of $N=8$ supergravity.
It is not immediately obvious that this is possible at all,
since the extra rotation must transform the family triplets $\bD_f$ and anti-triplets $\bbD_f$ with {\em opposite} phases, and it is {\em a priori} unclear whether and how
such a  transformation could be realized on the original 56 fermions of $N\!=\!8$
supergravity. Furthermore, enlarging SO(8) to the chiral SU(8) cannot help, as we
know that the U(1) that is associated with the electric charges must be vectorlike.

First we write out the correspondence more explicitly
\bea
\chi^{\bfa \bbE \bV} &\equiv & u^\bfa \;, \quad \chi^{\bfa \bbZ \bV} \equiv c^\bfa \; , \quad
\chi^{\bfa \bbD \bV} \equiv  t^\bfa  \nn \\
\chi^{\bfa \bbZ \bbD} &\equiv & d^\bfa \;, \quad \chi^{\bfa \bbD \bbE} \equiv s^\bfa \; , \quad
\chi^{\bfa \bbE \bbZ} \equiv  b^\bfa  \nn \\
\chi^{\bZ\bD\bbV} &\equiv & \nu_e \;, \quad \chi^{\bD \bE \bbV} \equiv \nu_\mu \; , \quad
\chi^{\bE \bZ\bbV} \equiv  \nu_\tau  \nn \\
\chi^{\bbZ \bbD \bbV} &\equiv & e^- \;, \quad \chi^{\bbD \bbE \bbV} \equiv \mu^- \; , \quad
\chi^{\bbE \bbZ\bbV} \equiv  \tau^-
\eea
where the boldface index $\bfa$ is the SU(3)$_c$ index (but remember that the
diagonal SU(3) rotates {\em all} indices different from $\bV$ and $\bbV$), and where
we ignore possible subtleties concerning the proper mass eigenstates, in particular
possible mixing with the Goldstino and `would-be Goldstinos' representations
in (\ref{Goldstino}) and (\ref{PseudoGoldstino}). Idem for the complex conjugate
representations which describe the associated anti-particles.
Hence, the searched for U(1)$_q$ rotation must act as follows
{\small
\begin{align}\label{U(1)q}
\delta u^\bfa &= - \,i \, u^\bfa &, \quad
\delta c^\bfa &= - \,i \, c^\bfa &, \quad
\delta t^\bfa &= - \,i \, t^\bfa  \nn\\
\delta d^\bfa &= + \,i \, d^\bfa &, \quad
\delta s^\bfa &= + \, i \, s^\bfa &, \quad
\delta b^\bfa &= + \, i \, b^\bfa  \nn\\
\delta \nu_e &= - \, i \, \nu_e &, \quad
\delta \nu_\mu &= - \, i \, \nu_\mu &, \quad
\delta \nu_\tau &= - \, i \, \nu_\tau   \nn\\
\delta e^- &=  + \, i \, e^- &, \quad
\delta \mu^- &=  + \, i \, \mu^- &, \quad
\delta \tau^- &=  + \, i \, \tau^- \; .
\end{align}
}
To find out whether and how this transformation can be realized on the
original spin-$\frac12$ fermions of the theory, we express the latter in terms of
the physical fermions, then perform the desired U(1)$_q$ rotation, and finally
transform back to the original basis. Although the intermediate expressions are
quite messy, the final result takes a very simple form. To this aim, consider the (vector-like) SO(8) generator (same as (\ref{U(1)}) with $\al = \be =1$)
\begin{equation} \label{q1}
 T \,=\, Y(1,1) \,
        = \, \begin{pmatrix}
                0 & 1  & 0 & 0 & 0 & 0 & 0 & 0 \\
                -1  & 0 & 0 & 0 & 0 & 0 & 0 & 0 \\
                0 & 0 & 0 & 1 & 0 & 0 & 0 & 0 \\
                0 & 0 & -1 & 0 & 0 & 0 & 0 & 0 \\
                0 & 0 & 0 & 0 & 0 &  1 & 0 & 0 \\
                0 & 0 & 0 & 0 & -1 & 0 & 0 & 0 \\
                0 & 0 & 0 & 0 & 0 & 0 & 0 & 1\\
                0 & 0 & 0 & 0 & 0 & 0 & -1 & 0
                \end{pmatrix},
\end{equation}
Next, introduce the following 56-by-56 matrix acting on the antisymmetrized
product of three $\bf{8}$ representations
\beq\label{cI}
\cI := \frac12 \Big(T \wedge {\bf 1} \wedge {\bf 1} \, + \,  {\bf 1} \wedge T \wedge {\bf 1} \, + \,
        {\bf 1} \wedge {\bf 1} \wedge T \, + \,  T \wedge T \wedge T \Big)
\eeq
Note that this is {\em not} the direct co-product that one would expect from (\ref{Uchi})
with $U = \exp (\omega T)$ acting on each of the three indices, and thus not even
an element of SU(8). Indeed, the extra term is reminiscent of the modification (`twist') required to deform a trivial into a non-trivial co-product. We note that, from $T^2 = -1$,
\beq
\cI^2 \, = \, - {\bf 1}\!\!{\bf 1}
\eeq
with the 56-by-56 unit matrix $\bf{1\!\!{\bf 1}}$,
which shows that (\ref{cI}) can be trivially exponentiated to a U(1)$_q$ phase rotation.
Examples of the action of $\cI$ are
\bea
\chi^{137} \;\; &\rightarrow& \quad  \frac12 \big(+ \chi^{237}  +  \chi^{147} +
   \chi^{138}  +  \chi^{248} \big) \nn\\
\chi^{247} \;\; &\rightarrow& \quad  \frac12 \big(- \chi^{147} - \chi^{237}
    + \chi^{248}   +  \chi^{138} \big)  \nn\\
%238 \;\; &\rightarrow& \quad  (-237 \; + \; 147 \: - \; 138 \; + \; 248)/2 \nn\\[1mm]
%148 \;\; &\rightarrow& \quad  (- 237 \; + \; 147 \: - \; 138 \; + \; 248)/2 \nn\\[1mm]
%237 \;\; &\rightarrow& \quad  (- 137 \; + \; 247 \: + \; 238 \; - \; 148)/2 \nn\\[1mm]
%147 \;\; &\rightarrow& \quad  (- 137 \; + \; 247 \: - \; 238 \; + \; 148)/2 \nn\\[1mm]
%138 \;\; &\rightarrow& \quad  (- 137 \; - \; 247 \: + \; 238 \; + \; 148)/2 \nn\\[1mm]
%248 \;\; &\rightarrow& \quad  (- 137 \; - \; 247 \: - \; 238 \; - \; 148)/2
\chi^{125} \;\; &\rightarrow& \quad \chi^{126} \quad , \quad \:
%\chi^{126} \;\; \rightarrow - \chi^{125} \nn\\[2mm]
%\chi^{345} \;\; &\rightarrow& \quad \chi^{346} \quad , \quad \:
\chi^{346} \;\; \rightarrow - \, \chi^{345} \;\; ,
\eea
and so on. While the commutation of $T\wedge T \wedge T$ with an arbitrary
element of SO(8) or SU(8) in the $\bf{56}$ representation would enlarge
either Lie algebra to a bigger one,
this is not the case for the residual SU(3)$\,\times\,$U(1) because $T$
(representing the imaginary unit) commutes with this subgroup. Hence,
it results in a genuine {\em deformation}, not an enlargement, of the residual
SU(3)$\,\times\,$U(1) symmetry at the stationary point.

Our main observation now is that $\cI$ does realize (\ref{U(1)q}), namely
the transformation
\beq
\chi^{ijk} \; \ra \quad \, \big(\cI \circ \chi)^{ijk}
\eeq
yields precisely the phase rotations shown in (\ref{U(1)q}), as is most
easily verified by observing that the phase is negative on $\chi$'s with
no or one barred index, and positive on $\chi$'s with two or three barred
indices (without the `twist term' in (\ref{cI}), the fermions would transform with a
phase factor $\exp (inq)$, where $n$ counts the number of barred
minus unbarred indices). Therefore, assigning all fermions the charge
$q=\frac16$ under U(1)$_q$ and combining the action of U(1)$_q$
with that of the supergravity U(1), we obtain the correct electric charges
for all 48 quarks and leptons. We emphasize that the simple formula
(\ref{cI}) appears to work only with the choice (\ref{albe}).

An obvious question at this point concerns the possible implementation,
within the present scheme, of chiral gauge interactions corresponding to the
full electroweak SU(2)$_w\,\times\,$U(1)$_Y$ symmetry (which our new U(1)$_q$
would be part of). While we have so far no definite answer to this question,
we would like to emphasize the following point. As already shown in \cite{dWN},
the gauging can be done while maintaining the `composite'  local SU(8) of \cite{CJ}.
In that formulation the theory has a local SO(8)$\,\times\,$SU(8) symmetry \cite{dWN},
which might play a role in explaining the emergence of chirality.
Indeed, when embedding the SU(3) subgroup of SO(8) into the (chiral) SU(8),
the maximal  symmetry that commutes with it is a (chiral) U(2) \cite{NW},
and this statement remains true with the deformed U(1) identified here.
While it is evident already from the discussion in \cite{NW} that this U(2)
by itself cannot produce the correct electroweak charge assignments,
because SU(2)$_w$ would not commute with SU(3)$_f$, a `twist' similar to
the one introduced here may be required to make things work. What is
clear, however, is that in such a scheme the $W^\pm$ and $Z$ vector bosons
would have to be {\em composite}, in a partial realization of the conjecture already
made in \cite{CJ}, that SU(8) becomes dynamical. We recall that the `composite'
chiral SU(8) symmetry does not suffer from anomalies \cite{Marcus}, and the same
should be true for any subgroup of SU(8) that becomes dynamical.

The results of this article lend further credence to the remarkable coincidence,
already exhibited in  \cite{GellMann} and \cite{NW}, between the fermionic
sector of $N\!=\!8$ supergravity  and the observed 48 spin-$\frac12$
fermions of the Standard Model. Evidently this agreement would be spoilt if any new
fundamental spin-$\frac12$ degrees of freedom (as predicted by all
models of $N\!=\!1$ low energy supersymmetry) were to be found at LHC.
While the numerology is thus very suggestive, there remain, of course, the
thorny open problems already listed in \cite{NW} (huge negative cosmological
constant, mass spectrum, {\em etc.}), whose resolution would demand some
new, and as yet unknown, dynamics which would also have to account
for the final breaking of $N\!=\!2$ supersymmetry. So the above coincidence
between theory and observation may yet  turn out to be a mirage. At any rate,
and in view of the complete absence so far of any `new physics' at LHC,
it appears worthwhile to search for unconventional alternatives, of the type
considered here, to currently popular
ideas.  In particular, the actual realization of supersymmetry in particle physics
may require a more sophisticated implementation of this beautiful concept
than in the $N\!=\!1$ models currently thought to be phenomenologically
viable.

\vspace{2mm}
\noindent
 {\bf Acknowledgments:} K.A.~Meissner thanks AEI for hospitality and support, and
H.~Nicolai thanks H.~Samtleben and ENS Lyon for hospitality, where this work
was supported by the Gay-Lussac-Humboldt Prize. We are also grateful to G. Dall'Agata
for comments on a first version of this paper.


\begin{thebibliography}{99}
\bibitem{fn} We might note that the existence of three families of 16 fermions each, for which there is now ample evidence, had not been fully established when \cite{GellMann} and \cite{NW} appeared.
\bibitem{Min} There do exist gaugings with Minkowski vacua and fully broken
supersymmetry which allow for other values of $\alpha$ and $\beta$ \cite{CDIZ},
but the choice (\ref{albe}) appears to be unique for the present scheme and the
simple formula (\ref{cI}) to work.
\end{thebibliography}
\end{document}